\documentclass[twocolumn,showpacs,showkeys,preprintnumbers,prd,superscriptaddress,nofootinbib]{revtex4-1}

\usepackage{amsmath}
\usepackage[]{hyperref}
\usepackage{cleveref}
\usepackage{graphicx}

\Crefname{equation}{Eq.}{Eqs.}
\Crefname{figure}{Fig.}{Figs.}
\Crefname{section}{Sec.}{Secs.}
\Crefrangeformat{equation}{Eqs.~(#3#1#4)--(#5#2#6)}

\begin{document}

\title{Model-independent reconstruction of  $f(Q)$ non-metric gravity}

\author{Salvatore Capozziello}
\email{capozziello@na.infn.it}
\affiliation{Dipartimento di Fisica ``E. Pancini", Universit\`a di Napoli ``Federico II", Via Cinthia 9, I-80126 Napoli, Italy.}
\affiliation{Scuola Superiore Meridionale, Largo S. Marcellino 10, I-80138 Napoli, Italy.}
\affiliation{Istituto Nazionale di Fisica Nucleare (INFN), Sez. di Napoli, Via Cinthia 9, I-80126 Napoli, Italy.}

\author{Rocco D'Agostino}
\email{rocco.dagostino@unina.it}
\affiliation{Scuola Superiore Meridionale, Largo S. Marcellino 10, I-80138 Napoli, Italy.}
\affiliation{Istituto Nazionale di Fisica Nucleare (INFN), Sez. di Napoli, Via Cinthia 9, I-80126 Napoli, Italy.}

\begin{abstract}
We consider gravity mediated by non-metricity, with vanishing curvature and torsion. The gravitational action, including an arbitrary function of  the non-metric scalar, is investigated in view of characterizing the dark energy effects. In particular, we present a method to reconstruct the $f(Q)$ action without resorting to \emph{a priori} assumptions on the cosmological model. To this purpose, we adopt a method based on rational Pad\'e approximations, which provides a stable behaviour of the cosmographic series at high redshifts, alleviating the convergence issues proper of the standard approach. We thus describe how to reconstruct $f(Q)$ through a numerical inversion procedure based on the current observational bounds on  cosmographic parameters. Our analysis suggests that the best approximation for describing the accelerated expansion of the universe is represented by a scenario with $f(Q)=\alpha+\beta Q^{n}$. Finally, possible deviations from the standard $\Lambda$CDM model are discussed.

\date{\today}

\end{abstract}

\pacs{04.50.Kd, 98.80.-k,  98.80.Es}

\maketitle

\section{Introduction}

Several observational evidences  support  the standard model of cosmology, the so-called $\Lambda$CDM model \cite{SN,BAO,Planck,mio_vittorio}. They provide the picture of a universe which  recently entered a phase of accelerated expansion. In the framework of general relativity (GR), the simplest interpretation of  dark energy, responsible for the late-time acceleration, is offered by the cosmological constant $\Lambda$, with a negative constant equation of state \cite{lambda}.

Despite the great success of $\Lambda$CDM to explain current observations, the interpretation of $\Lambda$ at fundamental level is still far from being understood, mainly due to the \emph{fine-tuning problem} inherent the energy of the vacuum \cite{fine-tuning}.
Among all scenarios attempting for an alternative explanation of  dark energy effects (see e.g. \cite{rocco1,rocco2}), particular attention in the last years has been devoted to modified gravity theories, including $f(R)$ and $f(T)$ theories \cite{review,f(R),f(T),rocco3}.

An interesting possibility explored very recently is to consider the gravitational interaction mediated by the non-metricity, while curvature and torsion are vanishing \cite{Jimenez,f(Q)_1,f(Q)_2,Sahoo}. This approach can be extremely important to describe gravity at fundamental level because gravity can be dealt as a gauge theory not requiring a priori the validity of the Equivalence Principle. 

 In this context, investigating the $f(Q)$ theories, where $Q$ is the non-metricity scalar,  could offer new insights on the cosmic speed up deriving from the intrinsic implications of a different geometry with respect to the Riemannian one.

Standard cosmological approaches to extended theories of gravity rely on assuming a specific form of the relative action, and then analyzing the cosmic behaviour to check for possible deviations from GR. This procedure, however, may induce an \emph{a priori} bias that could lead to misleading conclusions. 

These considerations motivated to investigate new model-independent techniques to tackle the dark energy problem. In such a context, the \emph{cosmographic approach} represents a powerful tool to study the kinematic features of the universe starting from first principles \cite{Weinberg72,cosmography1}. Indeed, the main advantages of cosmography reside in the fact that it does not need the assumption of a specific underlying cosmology to describe the dark energy behaviour \cite{cosmography2}. Applied to modified gravity theories, this method provides the remarkable opportunity to test the validity of GR on cosmological scales and to deal with possible deviations from the Einstein theory \cite{cosmography_extended}. 

In the present paper, we intend to apply the cosmographic method in order to obtain a model-independent reconstruction of $f(Q)$ gravity. After this Introduction, \Cref{sec:f(Q)} gives a brief summary of non-metricity gravity and $f(Q)$ cosmology. In \Cref{sec:cosmography}, we start by recalling the main ingredients of the cosmographic technique and, subsequently, we illustrate the numerical procedure to reconstruct the $f(Q)$ action based on the method of rational Pad\'e approximations. Finally, \Cref{sec:conclusions} is dedicated to a general discussion of our results and concluding remarks. 

Throughout this paper, we adopt natural units  $c=8\pi G=1$.

\section{Cosmology in $f(Q)$ gravity}
\label{sec:f(Q)}

In order to explore the cosmological features of non-metric gravity, let us consider the most general form of the affine connections \cite{Jarv18}:
\begin{equation}
\Gamma^{\lambda}_{\phantom{\alpha}\mu\nu} =
\left\lbrace {}^{\lambda}_{\phantom{\alpha}\mu\nu} \right\rbrace +
K^{\lambda}_{\phantom{\alpha}\mu\nu}+
 L^{\lambda}_{\phantom{\alpha}\mu\nu} \,.
\end{equation}
Here, the Levi-Civita connection of the metric tensor $g_{\mu\nu}$ is given as
\begin{equation}
\left\lbrace {}^{\lambda}_{\phantom{\alpha}\mu\nu} \right\rbrace \equiv \dfrac{1}{2}\,g^{\lambda \beta} \left( \partial_{\mu} g_{\beta\nu} + \partial_{\nu} g_{\beta\mu} - \partial_{\beta} g_{\mu\nu} \right),
\end{equation} 
whereas 
\begin{align}
K^{\lambda}{}_{\mu\nu}& \equiv \frac{1}{2}\, g^{\lambda \beta} \left(\mathcal T_{\mu\beta\nu}+\mathcal T_{\nu\beta\mu} +\mathcal T_{\beta\mu\nu} \right), \\
L^{\lambda}{}_{\mu\nu}& \equiv \frac{1}{2}\, g^{\lambda \beta} \left( -Q_{\mu \beta\nu}-Q_{\nu \beta\mu}+Q_{\beta \mu \nu} \right)
\end{align} 
are the contortion and disformation tensors, respectively. In the above definitions, we have also introduced the torsion tensor, $\mathcal T^{\lambda}{}_{\mu\nu}\equiv \Gamma^{\lambda}{}_{\mu\nu}-\Gamma^{\lambda}{}_{\nu\mu}$, and the non-metricity tensor, given by
\begin{equation}
Q_{\rho \mu \nu} \equiv \nabla_{\rho} g_{\mu\nu} = \partial_\rho g_{\mu\nu} - \Gamma^\beta{}_{\rho \mu} g_{\beta \nu} -  \Gamma^\beta{}_{\rho \nu} g_{\mu \beta}  \,.
\end{equation}
Thus, particular choices on the connections will specify the metric-affine spacetime. n our analysis, we assume that both curvature and torsion are vanishing, so that geometry is given by non-metricity. The non-metricity tensor is characterized by two independent traces, namely $Q_\mu=Q_{\mu\phantom{\alpha}\alpha}^{\phantom{\alpha}\alpha}$ and $\tilde{Q}^\mu={Q_{\alpha}}^{\mu\alpha}$ depending on the order of contraction. Hence, one can define the non-metricity scalar as \cite{Jimenez}
\begin{equation}
Q=-\dfrac{1}{4}Q_{\alpha\beta\mu}Q^{\alpha\beta\mu}+\dfrac{1}{2}Q_{\alpha\beta\mu}Q^{\beta\mu\alpha}+\dfrac{1}{4}Q_{\alpha}Q^{\alpha}-\dfrac{1}{2}Q_{\alpha}\tilde{Q}^\alpha\,,
\end{equation} 
which is a quadratic combination, invariant under general diffeomorphisms. 

In analogy to studies on torsionless $f(R)$ gravity and curvature-free $f(T)$ gravity, we can generalize the $Q$-gravity to theories containing an arbitrary function of the non-metricity scalar, i.e. $f(Q)$.
Therefore, we consider the following action:
\begin{equation}
S=\int  d^4x\, \sqrt{-g}\left[\dfrac{1}{2}f(Q)+\mathcal{L}_m\right],
\label{action}
\end{equation}
where $g$ is the determinant of the metric, and $\mathcal{L}_m$ is the Lagrangian density of the matter sector.

We note that, for $f(Q)=Q$, the above action is equivalent to the Einstein-Hilbert action up to a total derivative. In the case of a globally vanishing affine connections, the non-metricity tensor depends on the metric only and Einstein's GR action is recovered. This occurs under the choice of the \emph{coincidence gauge}, in which the origin of spacetime  and that of the tangent space coincide (we refer the reader to \cite{Beltran19} for the details).

The gravitational field equations are then obtained by varying the action with respect to the metric, leading to
\begin{widetext}
\begin{align}
&\dfrac{2}{\sqrt{-g}}\nabla_\alpha\bigg\{\sqrt{-g}\, g_{\beta\nu}\, f_Q\Big[-\dfrac{1}{2}L^{\alpha\mu\beta}- \dfrac{1}{8}\left(g^{\alpha\mu}Q^\beta+g^{\alpha\beta}Q^\mu\right)+\dfrac{1}{4}g^{\mu\beta}(Q^\alpha-\tilde{Q}^\alpha)\Big]\bigg\} \nonumber \\
&+f_Q\Big[-\dfrac{1}{2}L^{\mu\alpha\beta}-\dfrac{1}{8}\left(g^{\mu\alpha}Q^\beta+g^{\mu\beta}Q^\alpha\right)+\dfrac{1}{4}g^{\alpha\beta}(Q^\mu-\tilde{Q}^\mu)\Big]Q_{\nu\alpha\beta}+\dfrac{1}{2}{\delta^\mu}_\nu f={T^\mu}_\nu\,,
\label{eq:FE}
\end{align}
\end{widetext}
where $f_Q\equiv \partial f/\partial Q$ and $T_{\mu\nu}$ is the energy-momentum tensor, defined as
\begin{equation}
T_{\mu	\nu}=-\dfrac{2}{\sqrt{-g}}\dfrac{\delta \sqrt{-g} \mathcal{L}_m}{\delta g^{\mu\nu}}\,.
\end{equation}
For cosmological purposes, we consider the line element $ds^2=-dt^2+a(t)^2\delta_{ij}dx^idx^j$, corresponding to the spatially flat Friedman-Lema\^itre-Robertson-Walker (FLRW) metric, in which $a(t)$ is the cosmic scale factor used to define the Hubble expansion rate $H\equiv \dot{a}/a$.

Focusing our attention on the coincidence gauge, we can write the modified Friedman equations as \cite{Beltran20}
\begin{align}
6H^2f_Q-\dfrac{1}{2}f&=\rho\,, \label{eq:first Friedmann} \\
\left(12H^2f_{QQ}+f_Q\right)\dot{H}&=-\dfrac{1}{2}(\rho+p)\,,  \label{eq:second Friedmann}
\end{align}
where $\rho$ and $p$ are the total density and pressure of the cosmic fluid, respectively. 
Moreover, we have the following relation:
\begin{equation}
Q=6H^2\,,
\label{eq:Q-H}
\end{equation}
which will play a central role in our reconstruction procedure. 
It is worth to emphasize that \Cref{eq:first Friedmann,eq:second Friedmann,eq:Q-H} hold only in the \emph{coincidence gauge} or for the special case of connections fulfilling the spacetime symmetries \cite{D'Ambrosio22,Hohmann21}. In fact, choosing connections that vanish globally is not always possible in $f(Q)$ gravity, and one may end up with trivial solutions that cannot go beyond GR, regardless of the $f$ under consideration \cite{D'Ambrosio22bis}.

In our analysis, we focus on the late-time evolution of the cosmic fluid, so that we can neglect radiation and consider the entire contribution due to pressureless matter. This implies  $p=0$ and $\rho=3H_0^2 \Omega_{m0}(1+z)^{3}$, where the subscript zero refers to quantities evaluated at the present time, and $z$ is the redshift defined as $z\equiv a^{-1}-1$.\footnote{At the present time, $a(t_0)=1$ and $z=0$.}

\section{$f(Q)$ cosmography}
\label{sec:cosmography}

The cosmographic method allows to study the universe dynamics through kinematic quantities that do not depend on a specific background cosmology. Thus, by only assuming the homogeneity and isotropy of spacetime according to the Cosmological Principle, the late-time expansion history of the universe can be investigated in a model-independent way in order to extract information on the dark energy properties and the nature of gravity.

Before proceeding to the reconstruction of $f(Q)$ gravity, we briefly review in the following the main aspects of the cosmographic approach.

\subsection{The standard cosmographic recipe}

The key ingredient of cosmography is the Taylor expansion of the cosmological scale factor around the present time:
\begin{equation}
a(t)=1+\sum_{k=1}^{\infty}\dfrac{1}{k!}\dfrac{d^k a}{dt^k}\bigg | _{t=t_0}(t-t_0)^k\ .
\label{eq:scale factor}
\end{equation}
The above expansion defines the so-called \textit{cosmographic series}, whose first four terms are \cite{cosmography1}
\begin{subequations}
\begin{align}
&H(t)\equiv \dfrac{1}{a}\dfrac{da}{dt} \ , \hspace{1.14cm} q(t)\equiv -\dfrac{1}{aH^2}\dfrac{d^2a}{dt^2}\ ,  \label{eq:H&q} \\
&\hspace{0.16cm}j(t) \equiv \dfrac{1}{aH^3}\dfrac{d^3a}{dt^3} \ , \hspace{0.5cm}  s(t)\equiv\dfrac{1}{aH^4}\dfrac{d^4a}{dt^4}\ .    \label{eq:j&s}
\end{align}
\end{subequations}
These quantities are named, respectively,  Hubble, deceleration, jerk and snap parameters, and can be used to express cosmological distances without the need of an \emph{a priori} specific model.

Hence, inserting \Cref{eq:scale factor} into the definition of the luminosity distance, one readily finds
\begin{equation}
d_L(z)=\ \dfrac{z}{H_0}\left[1+d_L^{(1)}z+d_L^{(2)} z^2+d_L^{(3)}z^3+\mathcal{O}(z^4)\right],
\label{eq:luminosity distance}
\end{equation}
where 
\begin{subequations}
\begin{align}
d_L^{(1)}&=\dfrac{1}{2}(1-q_0)\,,		\\
d_L^{(2)}&=-\dfrac{1}{6}(1-q_0-3q_0^2+j_0)\,, \\
d_L^{(3)}&=\dfrac{1}{24}(2-2q_0-15q_0^2-15q_0^3+5j_0+10q_0j_0+s_0)\,.
\end{align}
\end{subequations}
Futhermore, using \Cref{eq:luminosity distance}, it is possible to parametrize the cosmic history starting from the relation
\begin{equation}
H(z)=\left[\dfrac{d}{dz}\left(\dfrac{d_L(z)}{1+z}\right)\right]^{-1}\,,
\label{eq:Hubble rate}
\end{equation}
which thus gives
\begin{equation}
H(z)= H_0\left[1+H^{(1)}z+H^{(2)}\dfrac{z^2}{2}+H^{(3)}\dfrac{z^3}{6}\right]+\mathcal{O}(z^4)\,,
\end{equation}
where
\begin{subequations}
\begin{align}
H^{(1)}&=1+q_0\,,\\
H^{(2)}&=j_0-q_0^2\,,\\
H^{(3)}&=3q_0^2+3q_0^3-j_0(3+4q_0)-s_0\,.
\label{eq:Taylor H(z)}
\end{align}
\end{subequations}

Although straightforward to implement,  the standard cosmographic method presents severe limitations when dealing with high-redshift, due to the short convergence radius proper of Taylor series. One possible way to alleviate such problem is to consider rational polynomials which are able to extend the convergence of the cosmographic series towards $z>1$. A relevant example in this respect is offered by Pad\'e approximations \cite{Pade}.

\subsection{High-redshift cosmography with Pad\'e approximations}

One of the most reliable cosmographic methods guaranteeing a stable behaviour at high redshifts is based on Pad\'e approximations. This is constructed starting from the Taylor series of a given function of the redshift, $f(z)=\sum_{k=0}^\infty c_kz^k$, where $c_k=f^{(k)}(0)/{k!}$ are constant coefficients. Thus, we can define the $(n,m)$ Pad{\'e} approximation of $f(z)$ as \cite{Litvinov93,Baker96}
\begin{equation}
P_{n, m}(z)=\dfrac{\displaystyle{\sum_{i=0}^{n}a_{i} z^{i}}}{\displaystyle{\sum_{j=0}^{m}b_j z^{j}}}\,.
\end{equation}
Requiring that $f(z)-P_{n,m}(z)=\mathcal{O}(z^{n+m+1})$, then one determines the coefficients of the above expansion as
\begin{equation}
\left\{
\begin{aligned}
&a_i=\sum_{k=0}^i b_{i-k}\ c_{k} \,,  \\
&\sum_{j=1}^m b_j\ c_{n+k+j}=-b_0\ c_{n+k}\,, \hspace{0.5cm} k=1,\hdots, m \,.
\end{aligned}
\right .
\end{equation}
The issue of choosing the degrees of rational polynomials has been addressed in the recent study \cite{Capozziello20}. In particular, it is  shown that Pad\'e approximations with polynomials of the same order in the numerator and denominator are prone to induce numerical errors and, thus, inaccurate cosmographic outcomes. Moreover,  a fitting analysis to low and high-redshift observations confirmed that the most suitable Pad\'e approximation is provided by the (2,1) polynomial, which also gives an optimal statistical performance given the low number of free coefficients involved.

In view of the aforementioned considerations, we here make use of the (2,1) Pad\'e approximation to obtain a reliable cosmographic reconstruction of  $f(Q)$ gravity. 
Specifically, the (2,1) approximation of the luminosity distance reads
\begin{equation}
d_{2,1}(z)=\dfrac{1}{H_0}\bigg[\dfrac{z (6 (-1 + q_0) + (-5 - 2 j_0 + q_0 (8 + 3 q_0)) z)}{-2 (3 + z + j_0 z) + 2 q_0 (3 + z + 3 q_0 z)}\bigg],
\label{eq:dL}
\end{equation}
and the corresponding Hubble expansion rate is
\begin{equation}
H_{2,1}(z)=H_0\dfrac{\mathcal{N}(z;q_0,j_0)}{\mathcal{D}(z;q_0,j_0)}\,,
\label{eq:H21}
\end{equation}
where 
\begin{subequations}
\begin{align}
\mathcal{N}(z;q_0,j_0)\equiv\ &2 (1 + z)^2 (3 + z + j_0 z - 3 q_0^2 z - q_0 (3 + z))^2, \\
\mathcal{D}(z;q_0,j_0)\equiv\ & 18 + 6 (5 + 2 j_0) z + (14 + 7 j_0 + 2 j_0^2) z^2 + 9 q_0^4 z^2 \nonumber \\
&+ 18 q_0^3 z (1 + z) - 2 q_0 (6 + 5 z) (3 + (4 + j_0) z)  \nonumber \\
&+ q_0^2 (18 + 30 z + (17 - 9 j_0) z^2)\,.
\end{align}
\end{subequations}
At this point, we note that \Cref{eq:dL,eq:H21} depend on a set of three cosmographic parameters, namely $\{H_0,q_0,j_0\}$, whose values are found through a direct comparison with data. Accordingly, in the present analysis, we adopt the numerical results obtained in \cite{Capozziello20} by means of the Markov Chain Monte Carlo integration technique applied to the combined likelihood of Supernovae Ia \cite{Pantheon} and observational Hubble data \cite{OHD}. We report below the results at  $1\sigma$ confidence level:
\begin{subequations}
\begin{align}
h_0&=0.693\pm 0.002\,, \label{h0} \\
q_0&=-0.73\pm 0.13\,,  \label{q0} \\
j_0&=2.84^{+1.00}_{-1.23}\,, \label{j0}
\end{align}
\end{subequations}
where $h_0\equiv H_0/$(100 km s$^{-1}$Mpc$^{-1})$. It is worth  remarking that these values have been obtained in the context of a flat universe by fixing the present matter density parameter as $\Omega_{m0}=0.3$.

\subsection{Reconstruction of the $f(Q)$ action}

\begin{figure}
\includegraphics[width=3.3in]{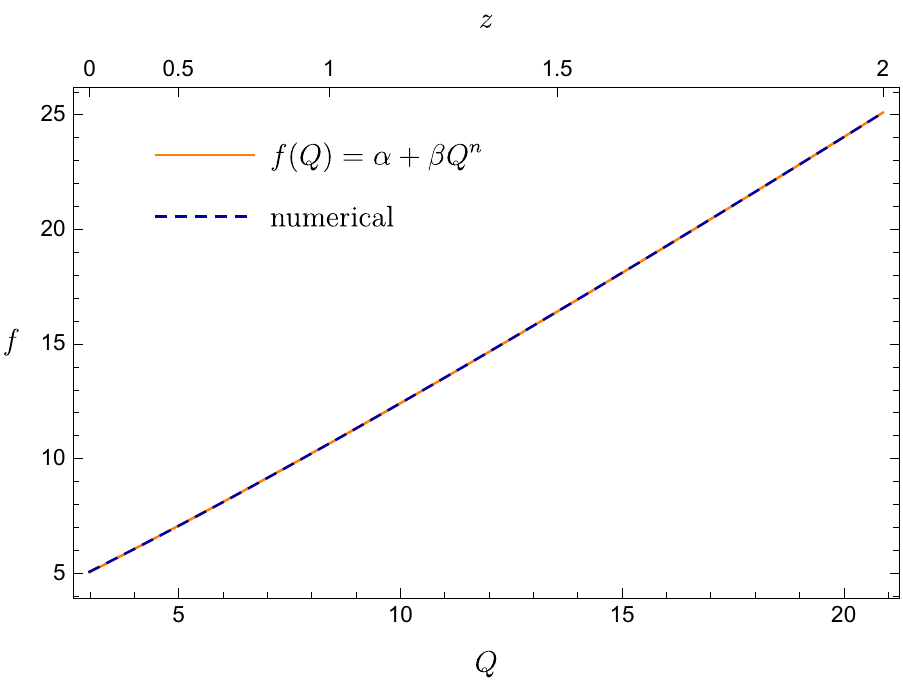}
\caption{Reconstruction of $f(Q)$ as a function of the redshift. The best analytical matching (solid orange) to the numerical solution (dashed blue) is provided by $f(Q)=\alpha+\beta Q^{n}$, with parameters values of \Cref{eq:best}.}
\label{fig:f(Q)}
\end{figure}

\begin{figure}
\includegraphics[width=3.3in]{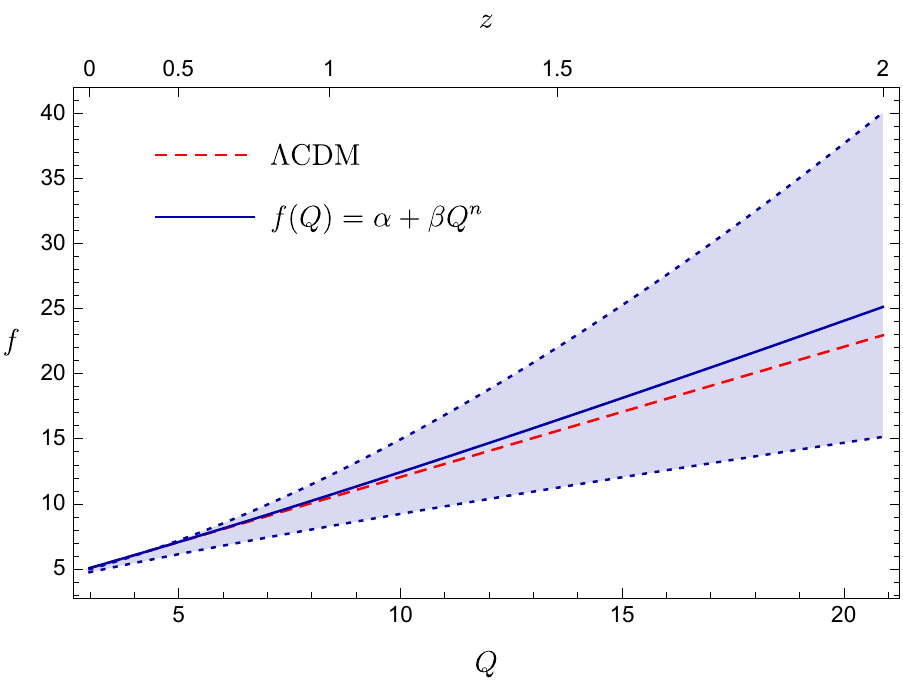}
\caption{Comparison between the reconstructed $f(Q)$ and $\Lambda$CDM.  The shaded regions around the best-fit curve of $f(Q)$ (cf. \Cref{eq:best}) take into account the lower and upper bounds at  $1\sigma$ confidence level (cf. \Cref{range_alpha,range_beta,range_n}). From the standard model, we assumed $\Omega_{m0}=0.3$ and $h_0=0.70$.}
\label{fig:comparison}
\end{figure}

To reconstruct the $f(Q)$ function of  the gravitational action, we consider the first Friedman equation \eqref{eq:first Friedmann}. Then, in view of \Cref{eq:Q-H}, we can convert the derivatives with respect to the non-metricity tensor in terms of derivatives of the Hubble parameter as a function of the redshift. Specifically, one finds
\begin{equation}
f_Q=\dfrac{f'(z)}{12\,H(z)\,H'(z)}	\,,
\label{eq:fQ}
\end{equation}
where the prime denotes derivative with respect to $z$. Using \Cref{eq:fQ}, we can recast \Cref{eq:first Friedmann} in the form
\begin{equation}
\dfrac{H'(z)}{H(z)}f'(z)-f(z)=6H_0^2\Omega_{m0}(1+z)^3\,.
\label{eq:master}
\end{equation}
Hence, assuming that the Hubble expansion is well approximated by \Cref{eq:H21}, and adopting the results given in \Cref{h0,q0,j0}, we can solve numerically \Cref{eq:master}.

In order to do that, a boundary condition is needed. To this purpose, we recall the effective gravitational constant in $f(Q)$ gravity, i.e. $G_\text{eff}\equiv G/f_Q$ \cite{Jimenez}. A reasonable  requirement is that $G_\text{eff}$ coincides with Newton's constant at  present epoch, which translates into having $f_Q=1$ at $z=0$. Applying this in \Cref{eq:first Friedmann} leads to the following initial condition:
\begin{equation}
f_0=6H_0^2(2-\Omega_{m0})\,.
\label{eq:boundary}
\end{equation}
Once $f(z)$ is known, we can  invert \Cref{eq:Q-H} by means of \Cref{eq:H21} and find $z(Q)$. Finally, the function $f(Q)$ is obtained by plugging $z(Q)$ back into $f(z)$. 
In so doing, we find that the numerical solution suitably matches with the function
\begin{equation}
f(Q)=\alpha+\beta Q^{n}\,,
\label{eq:test function}
\end{equation}
for the following set of constant coefficients:
\begin{equation}
(\alpha,\, \beta, \, n)=(2.492,\, 0.757, \, 1.118)\,.
\label{eq:best}
\end{equation}
We display our results in \Cref{fig:f(Q)}. We note that the test function \eqref{eq:test function} recovers pure GR for $\alpha=0$ and $\beta=1=n$, while the $\Lambda$CDM model for $\alpha >0$ and $\beta=1=n$. 

The numerical outcomes of \Cref{eq:best} suggest (small) deviations from the standard cosmological scenario. Possible inconsistencies with $\Lambda$CDM can be quantified by including, in the reconstruction procedure,  the 1$\sigma$ uncertainties over the cosmographic series as given in \Cref{h0,q0,j0}. Such an analysis yields
\begin{subequations}
\begin{align}
\alpha &\in[2.058,\, 3.162]\,, \label{range_alpha} \\
\beta & \in [0.332,\, 1.076]\,, \label{range_beta}\\
n&\in [0.821,\, 1.550]\,. \label{range_n}
\end{align}
\end{subequations}
We note that all the above ranges are consistent with the $\Lambda$CDM model. \Cref{fig:comparison} shows the reconstructed $f(Q)$ compared to the predictions of $\Lambda$CDM with $\Omega_{m0}=0.3$ and $h_0=0.70$. As one can see, the best-fit curves of the two scenarios are hardly distinguishable at late-times for $z< 1$, while they show small deviations from each other as the redshift increases. However, our results are well in agreement with the standard model within the $1\sigma$ confidence level.

\section{Discussion and conclusions}
\label{sec:conclusions}

Considering non-metricity as the mediator of gravitational interaction,  we focused on a gravitational action  containing a generic function of non-metricity scalar, which gives rise to the class of $f(Q)$ theories of gravity.

In this framework, the question of finding the $f(Q)$ able to provide the correct cosmological behaviour has been addressed in a model-independent way by means of cosmography. Relying only on the validity of the cosmological principle, such a method allows to reconstruct the $f(Q)$ gravity action without any \emph{a priori} ansatz on the underlying cosmological background. 

To do that, we performed a numerical reconstruction based  on rational Pad\'e approximations, which are able to reduce the convergence issues typical of the standard cosmographic technique, offering thus a reliable tool to describe cosmological observables up to high redshifts.
Taking into account constraints on the cosmographic series, obtained by a direct comparison with observations in the context of a flat universe, we expressed the Hubble expansion rate as a function of the redshift and, then, we exploited the relation $Q=6H^2$ to reconstruct $f(Q)$ through a numerical inversion procedure.

We found that the best analytical match to our numerical outcomes is given by the function $f(Q)=\alpha+\beta Q^n$, suggesting small deviations from the $\Lambda$CDM model as the redshift increases. However, including in the analysis the experimental uncertainties over the cosmographic parameters, our results indicate that departures from the standard cosmological model are not present at the $1\sigma$ confidence level.

It is worth noticing that the absence of significant deviations from $\Lambda$CDM is mainly due to the large uncertainty over the jerk parameter that propagates in our numerical analysis, limiting somewhat its effective predictability at very high redshifts. Nevertheless, this problem might be alleviated by the upcoming measurements from future experiments, which could provide more stringent bounds on higher-order terms of the cosmographic series and, thus, be more sensitive in testing possible inconsistencies with standard cosmology.

Furthermore, it is necessary to bear in mind the working hypothesis of the present study.  As pointed out in \Cref{sec:f(Q)}, our reconstruction procedure mainly relies on \Cref{eq:Q-H} that holds in particular cases, as the \emph{coincidence gauge}. Indeed, the same relation is not valid in other circumstances and this could lead to different results for the function $f(Q)$ \cite{D'Ambrosio22}.

Future efforts will be dedicated to compare the $f(Q)$ function here obtained with the large scale structure observations, to study the behaviour of our model at the perturbation level.
In this respect, the method presented here may represent a valuable tool to explore the intrinsic dark energy properties and break the degeneracy among cosmological models.

\acknowledgments 
S.C. and R.D. acknowledge the support of  Istituto Nazionale di Fisica Nucleare, Sez. di Napoli, {\it iniziativa specifiche} QGSKY and MoonLIGHT2.

{}

\end{document}